\documentclass[12pt]{article}
\usepackage{amsfonts,amssymb,amsmath}
\usepackage{setspace}

\begin{document}
\title{Weakly nonlocal thermodynamics \\of binary mixtures of Korteweg fluids\\
with two velocities and two temperatures}

\author{V.~A.~Cimmelli${}^{a}$, M.~Gorgone${}^{b}$, F.~Oliveri${}^{b}$, A.~R.~Pace${}^{a}$\\
\ \\
{\footnotesize ${}^a$ Department of Mathematics, Computer Science,}\\
{\footnotesize and Economics, University of Basilicata}\\
{\footnotesize Viale dell'Ateneo Lucano 10, 85100, Potenza, Italy}\\
{\footnotesize vito.cimmelli@unibas.it; raffaele.pace@unibas.it}\\
{\footnotesize ${}^b$ Department of Mathematical and Computer Sciences,}\\
{\footnotesize Physical Sciences and Earth Sciences, University of Messina}\\
{\footnotesize Viale F. Stagno d'Alcontres 31, 98166 Messina, Italy}\\
{\footnotesize mgorgone@unime.it; foliveri@unime.it}
}

\date{Published in \textit{European J. Mech./B Fluids} \textbf{83}, 58--65 (2020).}

\maketitle

\begin{abstract}
We provide a thermodynamic framework for binary mixtures of Korteweg fluids with two velocities and two 
temperatures. The constitutive functions are allowed to depend on the diffusion velocity and the specific internal 
energy of both constituents, together with their first gradients, as well as on the mass density of the mixture and the 
concentration of one of the constituents, the latters together with their first and second gradients. Compatibility with 
second law of thermodynamics is investigated by applying a generalized Liu procedure. 
In the one-dimensional case, a complete solution of the set of thermodynamic restrictions is obtained by postulating 
a possible form of the constitutive equations for the partial heat fluxes and stress tensors.  Taking a first order 
expansion in the gradients of the specific entropy, the expression of the entropy flux is determined. This contains the 
classical terms (namely, the sum of the ratios between the heat fluxes and the temperatures of the constituents) and 
some additional contributions accounting for  nonlocal effects.
\end{abstract}


%\MSC[2008]{74A15 \sep 74A20 \sep 76A10}

\section{Introduction}
\label{sec:intro}
Mixtures are a classical subject of rational thermodynamics \cite{Mul1,GurVar,Bow,LiuMul}, since they open a wide 
perspective on several basic problems of continuum physics \cite{TRUE}.

In the general case, the evolution of an $N$-component mixture is described by $5N$ fundamental fields, namely 
the $N$ partial mass densities $\rho^A$ $(A=1,\ldots, N)$, the $N$ partial velocities $\mathbf{v}^A$, and the $N$ 
partial temperatures $\theta^A$, (or, equivalently, internal energies $\varepsilon^{A}$) of the constituents.
With these basic fields we can derive some quantities related to the mixture as a whole (mass density, barycentric 
velocity and internal energy of the mixture).

Mixtures can be modeled at different degrees of detail (see, for instance, \cite{BowGar,PPR2005,GouRug,BotDre}), 
so that one can find:
\begin{enumerate}
\item models where the fundamental fields are the densities of each constituent, the barycentric velocity and the 
temperature of the mixture; 
\item models where the fundamental fields are the mass densities and the partial temperatures of each constituent, 
together with the barycentric velocity of the mixture; 
\item models where the fundamental fields are the mass densities and the velocities of each constituent, together 
with the temperature of the mixture; 
\item models where the fundamental fields are the mass densities, the partial temperatures, and the velocities of 
each constituent.
\end{enumerate}

It is worth noticing that in some physical problems involving mixtures of fluids, an internal state variable as an 
additional fundamental field  can be introduced \cite{FraPalRog,FraPalRog1,OliPalRog}. 
Nevertheless, here we will not consider such a situation.

Among the above models, the first one is the most classical \cite{GurVar,LiuMul}, and is related to the basic problem 
of the appropriate  form of the local balances of energy and entropy in continuum theories. 
In fact, in order to ensure the compatibility of some models of type 1 with second law of thermodynamics, some 
authors introduced an additional rate of supply of mechanical energy, the interstitial working, engendered by long-
range interactions among the molecules \cite{DunSer,Dun}, which results in an energy extra-flux 
${\mathbf{L}}$ into the local balance of energy.

Alternatively, it may be supposed that
the entropy flux $\mathbf{J}$ is affected by additional mechanical terms, due to long-range
forces, so that it is no longer assumed in the classical form $\mathbf{J}=\displaystyle\frac{\mathbf{q}}{\theta}$
\cite{Mul}, where ${\mathbf{q}}$ denotes the heat flux and $\theta$ is the absolute temperature. 
This is tantamount to introduce in the classical balance of entropy \cite{ColNol} an entropy
extra-flux $\mathbf{K}= \mathbf{J}-\displaystyle\frac{\mathbf{q}}{\theta}$.

It is worth of being remarked that these two options are not equivalent, and lead to
different expressions for the local entropy production (see \cite{CimOliPac3,CimSelTri3} for an extensive discussion).

The models of type 2 represent a more detailed description of the thermodynamics of the mixtures, which is relevant, 
for instance, in plasma theories, where the different constituents of the plasma may experience different 
temperatures on time scales of the same order of magnitude of the transport process times. This problem has been 
first considered in \cite{BowGar} and, more recently, in \cite{GouRug}.

The models of type 3 are important if one aims to obtain a more accurate description of the dynamics of the mixture, 
leaving the thermodynamic description at a classical level. Such a description may also provide more information on 
the constitutive quantities. Recently, models of type 3 have been considered in \cite{BotDre}.

To the best of our knowledge,  scant attention has been payed heretofore in the literature to models of type 4, which 
are, instead, the focus of our investigation.  In fact, in the present paper we consider a binary mixture for which the 
fundamental fields are the mass densities, the partial temperatures and the velocities of both the constituents.

The constituents of the mixture are assumed to be two Korteweg fluids, \emph{i.e.}, fluids whose state space is 
allowed to include the second spatial derivatives of the mass densities \cite{DunSer}. For this class of fluids the 
Cauchy stress tensor
has a constitutive equation \cite{Kor} like
\begin{equation} 
\label{kort}
T_{ij}=\left(-p+\sum_{k=1}^3\left(\alpha_1\frac{\partial^2\rho}{\partial x_k^2}+\alpha_2\frac{\partial\rho}{\partial x_k}
\frac{\partial\rho}{\partial x_k}\right)\right)\delta_{ij}+\alpha_3\frac{\partial\rho}{\partial x_i}
\frac{\partial\rho}{\partial x_j}+\alpha_4\frac{\partial^2\rho}{\partial x_i\partial x_j},
\end{equation}
where $\rho$ denotes the mass density, $p$ is the pressure of the fluid, and $\alpha_i$, $i=1,\ldots,4$,  
suitable material functions depending on density and temperature. These fluids received a moderate attention in the 
literature after the pioneering paper by Dunn and Serrin \cite{DunSer}, where the compatibility with the basic tenets 
of rational continuum thermodynamics \cite{TRUE} has been extensively studied. They have been studied also in 
\cite{CimOliPac3,CimSelTri3,CimSelTri1,CimOlTri} by means of a generalized Liu procedure \cite{Liu,Cim1}, and by 
Heida and M\'alek \cite{HeiMal} following a different methodology. 

The main results achieved in the present paper are the following ones.
First, we develop a consistent thermodynamic framework for binary mixtures of Korteweg fluids with two velocities 
and two temperatures, and prove that, for our model, second law of thermodynamics allows the dependence of the 
constitutive equations on all the gradients entering the state space.
Then, we prove that the thermodynamic restrictions placed by second law of thermodynamics are compatible with a 
very general form of the Cauchy stress tensor which encompasses the constitutive equation of Korteweg fluids.
Finally, in the one-dimensional case, we derive an explicit solution of the system of thermodynamic restrictions, so 
providing a representation for the specific entropy, the Cauchy stress tensor, the heat flux and the entropy flux.

The paper is organized as follows.
In Section~\ref{sec:balance}, we write the balance equations of mass, linear momentum and energy for both 
constituents, together with the unbalance of entropy, which expresses, locally, the second law of thermodynamics. 
Then, we illustrate the generalized procedure we apply for the exploitation of the entropy principle. 
In Section~\ref{sec:liu}, we derive a set of conditions ensuring that second law of thermodynamics is satisfied for 
arbitrary thermodynamic processes.
In Section~\ref{sec:1d}, we consider a one-dimensional system and provide a solution of the restrictions obtained in 
Section~\ref{sec:liu} by determining explicit constitutive equations for Cauchy stress tensor, heat flux, entropy and 
entropy flux. The lengthy computations necessary to derive and solve the compatibility restrictions placed by second 
law of thermodynamics are handled with the help of a package written by one of the authors (F.O.) in the Computer 
Algebra System Reduce \cite{Reduce}. 

A discussion about the physical meaning of the particular solution derived in Section~\ref{sec:1d}, and  possible 
developments of the present theory, is carried out in Section~\ref{sec:final}.

\section{Balance equations, entropy inequality and generalized Liu procedure}
\label{sec:balance}
Let us consider a non-reacting binary mixture of two Korteweg fluids. In the absence of external forces and  heat 
sources, the balance equations for  mass, linear momentum and energy of each constituent are
\begin{equation}
\begin{aligned}
&\rho^{(A)}_{,t}+v^{(A)}_{j}\rho^{(A)}_{,j}+\rho^{(A)} v^{(A)}_{j,j}=0,\\
&\rho^{(A)}(v^{(A)}_{i,t}+v^{(A)}_{j}v^{(A)}_{i,j})-T^{(A)}_{ij,j}=0,\\
&\rho^{(A)}(\varepsilon^{(A)}_{,t}+v^{(A)}_j\varepsilon^{(A)}_{,j})-T^{(A)}_{ij}v^{(A)}_{i,j}+q^{(A)}_{j,j}=0,
\end{aligned}
\end{equation}
where the upscript ${}^{(A)}$ ($A=1,2$) labels the two constituents, $\rho^A$ denotes the mass density,
$v^{(A)}_i$ the components of the velocity, $\varepsilon^{(A)}$ the internal energy per unit volume, 
$T^{(A)}_{ij}$ the components of the Cauchy stress tensor, and $q^{(A)}_i$ the components of the heat flux. 
Moreover, the subscripts
${}_{,t}$ and ${}_{,j}$ denote the partial derivatives with respect to $t$ and $x_j$, respectively, and the Einstein 
convention on sums over repeated indices has been used. 

It is worth observing that, since the velocities of the fluid components are different, the convective time derivatives 
for the two fluids are not the same.

From now on, let us describe the mixture by using as fields the density $\rho$ of the whole mixture together with the 
concentration $c$ of the first constituent  \cite{GurVar,LiuMul},  \emph{i.e.}, 
\begin{equation}
\rho=\rho^{(1)}+\rho^{(2)}, \qquad c=\frac{\rho^{(1)}}{\rho},
\end{equation}
the velocity $\mathbf{v}$ of the whole mixture and the diffusion velocity $\mathbf{w}$ of the first constituent, 
\emph{i.e.},
\begin{equation}
\mathbf{v}= \frac{\rho^{(1)} \mathbf{v}^{(1)}+\rho^{(2)} \mathbf{v}^{(2)}}{\rho}, \qquad 
\mathbf{w}=\mathbf{v}^{(1)}-\mathbf{v}.
\end{equation}
Finally, as far as the balance of energy is concerned, we take the equations for each constituent.

Therefore, the governing equations of our mixture become
\begin{equation}
\label{modelequations}
\begin{aligned}
\mathcal{E}^{(1)}\equiv&\rho_{,t}+v_{j}\rho_{,j}+\rho v_{j,j}=0,\\
\mathcal{E}^{(2)}\equiv&\rho(c_{,t}+v_j c_{,j})+\left(\rho c w_j\right)_{,j}=0,\\
\mathcal{E}^{(3)}_i\equiv&\rho(v_{i,t}+v_{j}v_{i,j})-\left(T^{(1)}_{ij}+T^{(2)}_{ij}-\frac{\rho c}{1-c}w_iw_j\right)_{,j}=0,\\
\mathcal{E}^{(4)}_i\equiv&\rho c\left(w_{i,t}+(v_j+w_j)w_{i,j}+w_j v_{i,j}\right)+(c-1)T^{(1)}_{ij,j}+cT^{(2)}_{ij,j}\\
&-c\left(\frac{\rho c}{1-c} w_iw_j\right)_{,j}=0,\\
\mathcal{E}^{(5)}\equiv&\rho c(\varepsilon^{(1)}_{,t}+(v_{j}+w_{j})\varepsilon^{(1)}_{,j})-T^{(1)}_{ij}(v_{i}+w_{i})_{,j}
+q^{(1)}_{j,j}=0,\\
\mathcal{E}^{(6)}\equiv&\rho (1-c)\left(\varepsilon^{(2)}_{,t}+
\left(v_{j}-\frac{cw_{j}}{1-c}\right)\varepsilon^{(2)}_{,j}\right)\\
&-T^{(2)}_{ij}\left(v_{i}-\frac{c}{1-c}w_{i}\right)_{,j}+q^{(2)}_{j,j}=0.
\end{aligned} 
\end{equation}

Finally, we have the local entropy inequality that, written for the whole mixture, takes the form
\begin{equation}
\label{entropyinequality}
\rho (s_{,t}+v_j s_{,j})+J_{j,j}\ge 0,
\end{equation}
where $s$ is the specific entropy, and $J_{j}$ are the components of the entropy flux.

The model equations and the entropy inequality, once the variables entering the state space are fixed, must be 
supplemented by the constitutive equations for the Cauchy stress tensors, the heat fluxes, the specific entropy and 
the entropy flux.

In view of equation~\eqref{kort},  let us assume the state space 
to be spanned by
\begin{equation}
\label{statespace}
\mathcal{Z}=\{\rho,c,w_i,\varepsilon^{(1)},\varepsilon^{(2)},\rho_{,j},c_{,j},v_{i,j},w_{i,j},\varepsilon^{(1)}_{,j},
\varepsilon^{(2)}_{,j},\rho_{,jk},c_{,jk}\},
\end{equation}
\emph{i.e.}, let us allow the constitutive equations to depend also on the second order spatial derivatives of $\rho$ 
and $c$; of course, the velocity gradient must enter the state space through its symmetric part.

The entropy principle imposes that the inequality \eqref{entropyinequality} must be satisfied for arbitrary 
thermodynamic processes \cite{CimJouRugVan,JCL}. To find a set of conditions which are at least sufficient for the 
fulfillment of such a constraint, we apply a generalized Liu procedure recently developed in a series of papers 
\cite{CimSelTri1,CimOlTri,Cim1}, incorporating new restrictions consistent with higher order nonlocal constitutive 
theories.
According to the procedure developed in \cite{CimSelTri1,CimOlTri,Cim1}, in exploiting the second law we take into 
account the balance equations for the unknown fields, and their gradient extensions too, up to the order of the 
gradients entering the state space.

Simple mathematical considerations may clarify the necessity of imposing as additional constraints in the 
entropy inequality the gradients of the balance equations when dealing with nonlocal constitutive equations. 

The thermodynamic processes are solutions of the balance equations, and, if these solutions are smooth enough, 
are trivially solutions of their differential consequences  (see also \cite{RogCim}). 
Since the entropy inequality \eqref{entropyinequality} has to be satisfied in arbitrary smooth processes,  then it is 
natural, from a mathematical point of view, to use the differential consequences of the equations governing those 
processes as constraints for such an inequality. On the contrary, if we restrict ourselves to consider as constraints 
only the balance equations, we are led straightforwardly to a specific entropy and Lagrange multipliers which are 
independent of the gradients of the field variables entering the state space, as it is shown by means of a simple 
example at the end of Section~\ref{sec:liu}. As a direct consequence, a Cauchy stress tensor depending on the 
gradients of mass density should be incompatible with second law of thermodynamics. Last, but not the least, the
extended Liu procedure does not require neither to assume a modified energy equation including additional 
contributions (such as the interstitial work \cite{DunSer}) nor an \emph{ad hoc} extra-flux in the entropy inequality; 
remarkably, an entropy flux given by the classical term and possible additional contributions arises in a natural way 
from the procedure itself.

\section{Thermodynamic restrictions}
\label{sec:liu}
As we said in the previous section, in exploiting second law of thermodynamics we have to take into account the constraints 
imposed on the thermodynamic processes by the balance equations and their gradient extensions; this task is
accomplished by introducing suitable Lagrange multipliers. 
Therefore, the entropy inequality writes:
\begin{equation}
\label{entropyconstrained}
\begin{aligned}
&\rho (s_{,t}+v_j s_{,j})+J_{j,j} \\
&\quad- \lambda^{(1)} \mathcal{E}^{(1)}- \lambda^{(2)} \mathcal{E}^{(2)}- \lambda^{(3)}_i \mathcal{E}^{(3)}_i- \lambda^{(4)}_i \mathcal{E}^{(4)}_i- \lambda^{(5)} \mathcal{E}^{(5)}- \lambda^{(6)} \mathcal{E}^{(6)}\\
&\quad-\Lambda^{(1)}_k\mathcal{E}^{(1)}_{,k}-\Lambda^{(2)}_k\mathcal{E}^{(2)}_{,k}-\Lambda^{(3)}_{ik}\mathcal{E}^{(3)}_{i,k}-
\Lambda^{(4)}_{ik}\mathcal{E}^{(4)}_{i,k}-\Lambda^{(5)}_{k}\mathcal{E}^5_{,k}-\Lambda^{(6)}_{k}\mathcal{E}^{(6)}_{,k}\\
&\quad-\Lambda^{(1)}_{k\ell}\mathcal{E}^{(1)}_{,k\ell}-\Lambda^{(2)}_{k\ell}\mathcal{E}^{(2)}_{,k\ell} \geq0.
\end{aligned}
\end{equation}

Expanding the derivatives entering the inequality \eqref{entropyconstrained}, we obtain a very long expression
where we can distinguish the \emph{highest derivatives} and the \emph{higher derivatives}  \cite{CimSelTri1}. 
The highest derivatives are both the time derivatives of the field variables and of the
elements of the state space, which cannot be expressed through the governing equations as functions of the 
thermodynamic variables, and the spatial derivatives whose order is the highest one. On the contrary, the higher 
derivatives are the spatial derivatives whose order is not maximal but higher than that of the gradients entering the 
state space. In the classical case, the highest derivatives coincide with the higher ones. 
In the case under study, the highest derivatives are 
\begin{equation}
\begin{aligned}
\mathbf{X}=\{&\rho_{,t},c_{,t},v_{i,t},w_{i,t},\varepsilon^{(1)}_{,t},\varepsilon^{(2)}_{,t},\rho_{,kt},c_{,kt},v_{i,kt},w_{i,kt},
\varepsilon^{(1)}_{,kt},\varepsilon^{(2)}_{,kt},\\
&\rho_{,k\ell t},c_{,k\ell t},\rho_{,jk\ell m},c_{,jk\ell m},v_{i,jk\ell},
w_{i,jk\ell},\varepsilon^{(1)}_{,jk\ell},\varepsilon^{(2)}_{,jk\ell}\},
\end{aligned}
\end{equation}
whereas the higher derivatives are
\begin{equation}
\mathbf{Y}=\{\rho_{,jk\ell},c_{,jk\ell},v_{i,jk},w_{i,jk},\varepsilon^{(1)}_{,jk},\varepsilon^{(2)}_{,jk}\}.
\end{equation}

By extensive though straightforward computation, it is easily ascertained that \eqref{entropyconstrained} 
can be written in the compact form
\begin{equation}
\label{entropycompatta}
A_p X_p+B^3_{qrs}Y_qY_rY_s+B^2_{qr}Y_qY_r+B^1_qY_q+C\ge 0,
\end{equation}
where the functions $A_p$, $B^1_{q}$, $B^2_{qr}$, $B^3_{qrs}$ and $C$ depend on the field and state variables 
only;
this means that the entropy inequality is linear in the highest derivatives and cubic in the higher ones.
The value of the highest and higher derivatives can be arbitrarily assigned \cite{CimOlTri} 
independently of the value of $C$, which, instead, is defined on the state space. 
First of all, let us observe that in principle nothing prevents the possibility of 
a thermodynamic process where $C=0$. Moreover, since the inequality \eqref{entropycompatta} has to be satisfied 
for arbitrary $X_{p}$ and $Y_q$, 
the conditions 
\begin{equation}
A_p=0, \quad B^3_{qrs}=0, \quad B^1_{q}=0,\quad C\ge 0,
\end{equation} 
as well the requirement that the quantities $B^2_{qr}$ are the entries of a semidefinite positive symmetric matrix, are 
sufficient for the fulfillment of
the entropy inequality.
By imposing the coefficients of the highest derivatives to be vanishing, we get the constraints
\begin{equation}
\begin{aligned}
&\rho\frac{\partial s}{\partial\rho}-\lambda^{(1)}=0,\allowdisplaybreaks\\
&\rho\frac{\partial s}{\partial c}-\rho\lambda^{(2)}-\rho_{,k}\Lambda^{(2)}_k-\rho_{,jk}\Lambda^{(2)}_{jk}=0,
\allowdisplaybreaks\\
&\rho \lambda^{(3)}_i+\rho_{,k}\Lambda^{(3)}_{ik}=0,\allowdisplaybreaks\\
&\rho\frac{\partial s}{\partial w_i}-\rho c\lambda^{(4)}_i-(\rho c_{,k}+c\rho_{,k})\Lambda^{(4)}_{ik}=0,\\
&\rho\frac{\partial s}{\partial \varepsilon^{(1)}}-\rho c\lambda^{(5)}-(\rho c_{,k}+c\rho_{,k})\Lambda^{(5)}_{k}
=0,\allowdisplaybreaks\\
&\rho\frac{\partial s}{\partial \varepsilon^{(2)}}-\rho(1-c)\lambda^{(6)}-(-\rho c_{,k}+(1-c)\rho_{,k})\Lambda^{(6)}_{k}
=0,\allowdisplaybreaks\\
&\rho\frac{\partial s}{\partial\rho_{,k}}-\Lambda^{(1)}_k=0,\allowdisplaybreaks\\
&\rho\frac{\partial s}{\partial c_{,k}}-\rho\Lambda^{(2)}_k-2\rho_{,j}\Lambda^{(2)}_{jk}=0,\\
&\frac{\partial s}{\partial v_{i,k}}-\rho \Lambda^{(3)}_{ik}=0,\allowdisplaybreaks\\
&\rho\frac{\partial s}{\partial w_{i,k}}-\rho c\Lambda^{(4)}_{ik}=0,\allowdisplaybreaks\\
&\rho\frac{\partial s}{\partial \varepsilon^{(1)}_{,k}}-\rho c\Lambda^{(5)}_{k}=0,\allowdisplaybreaks\\
&\rho\frac{\partial s}{\partial \varepsilon^{(2)}_{,k}}-\rho (1-c)\Lambda^{(5)}_{k}=0,\allowdisplaybreaks\\
&\rho\frac{\partial s}{\partial\rho_{,jk}}-\Lambda^{(1)}_{jk}=0,\allowdisplaybreaks\\
&\rho\frac{\partial s}{\partial c_{,jk}}-\rho\Lambda^{(2)}_{jk}=0,
\end{aligned}
\end{equation}
whence we are able to determine the Lagrange multipliers,
as well as the restrictions
\begin{equation}
\begin{aligned}
&\left\langle\left(\Lambda^{(3)}_{ik}-(c-1)\Lambda^{(4)}_{ik}\right)\frac{\partial T^{(1)}_{ij}}{\partial v_{l,m}}
+\left(\Lambda^{(3)}_{ik}-c\Lambda^{(4)}_{ik}\right)\frac{\partial T^{(2)}_{ij}}{\partial v_{l,m}}\right.\\
&\qquad
\left.-\Lambda^{(5)}_{k}\frac{\partial q^{(1)}_{j}}{\partial v_{l,m}}-\Lambda^{(6)}_{k}\frac{\partial q^{(2)}_{j}}{\partial 
v_{l,m}}\right\rangle_{(jk\ell m)}=0,\allowdisplaybreaks\\
&\left\langle\left(\Lambda^{(3)}_{ik}-(c-1)\Lambda^{(4)}_{ik}\right)\frac{\partial T^{(1)}_{ij}}{\partial w_{l,m}}
+\left(\Lambda^{(3)}_{ik}-c\Lambda^{(4)}_{ik}\right)\frac{\partial T^{(2)}_{ij}}{\partial w_{l,m}}\right.\\
&\qquad\left.-\Lambda^{(5)}_{k}\frac{\partial q^{(1)}_{j}}{\partial w_{l,m}}-\Lambda^{(6)}_{k}\frac{\partial q^{(2)}_{j}}
{\partial w_{l,m}}\right\rangle_{(jk\ell)}=0,\allowdisplaybreaks\\
&\left\langle\left(\Lambda^{(3)}_{ik}-(c-1)\Lambda^{(4)}_{ik}\right)
\frac{\partial T^{(1)}_{ij}}{\partial \varepsilon^{(1)}_{,\ell}}
+\left(\Lambda^{(3)}_{ik}-c\Lambda^{(4)}_{ik}\right)\frac{\partial T^{(2)}_{ij}}{\partial \varepsilon^{(1)}_{,\ell}}\right.\\
&\qquad\left.-\Lambda^{(5)}_{k}\frac{\partial q^{(1)}_{j}}{\partial \varepsilon^{(1)}_{,\ell}}-\Lambda^{(6)}_{k}
\frac{\partial q^{(2)}_{j}}{\partial \varepsilon^{(1)}_{,\ell}}\right\rangle_{(jk\ell)}=0,\allowdisplaybreaks\\
&\left\langle\left(\Lambda^{(3)}_{ik}-(c-1)\Lambda^{(4)}_{ik}\right)
\frac{\partial T^{(1)}_{ij}}{\partial \varepsilon^{(2)}_{,\ell}}
+\left(\Lambda^{(3)}_{ik}-c\Lambda^{(4)}_{ik}\right)\frac{\partial T^{(2)}_{ij}}{\partial \varepsilon^{(2)}_{,\ell}}\right.\\
&\qquad\left.-\Lambda^{(5)}_{k}\frac{\partial q^{(1)}_{j}}{\partial \varepsilon^{(2)}_{,\ell}}-\Lambda^{(6)}_{k}
\frac{\partial q^{(2)}_{j}}{\partial \varepsilon^{(2)}_{,\ell}}\right\rangle_{(jk\ell)}=0,\allowdisplaybreaks\\
&\left\langle\left(\Lambda^{(3)}_{ik}-(c-1)\Lambda^{(4)}_{ik}\right)\frac{\partial T^{(1)}_{ij}}{\partial \rho_{,\ell m}}
+\left(\Lambda^{(3)}_{ik}-c\Lambda^{(4)}_{ik}\right)\frac{\partial T^{(2)}_{ij}}{\partial \rho_{,\ell m}}\right.\\
&\qquad\left.-\Lambda^{(5)}_{k}\frac{\partial q^{(1)}_{j}}{\partial \rho_{,\ell m}}-
\Lambda^{(6)}_{k}\frac{\partial q^{(2)}_{j}}{\partial \rho_{,\ell m}}\right\rangle_{(jk\ell m)}=0,\allowdisplaybreaks\\
&\left\langle\left(\Lambda^{(3)}_{ik}-(c-1)\Lambda^{(4)}_{ik}\right)\frac{\partial T^{(1)}_{ij}}{\partial c_{,\ell m}}
+\left(\Lambda^{(3)}_{ik}-c\Lambda^{(4)}_{ik}\right)\frac{\partial T^{(2)}_{ij}}{\partial c_{,\ell m}}\right.\\
&\qquad\left.-\Lambda^{(5)}_{k}\frac{\partial q^{(1)}_{j}}{\partial c_{,\ell m}}
-\Lambda^{(6)}_{k}\frac{\partial q^{(2)}_{j}}{\partial c_{,\ell m}}\right\rangle_{(jk\ell m)}=0,
\end{aligned}
\end{equation}
where the symbol $\langle\mathcal{T}\rangle_{(jk\ldots)}$ stands for the symmetric part of $\mathcal{T}$ with respect to the indices 
$j$, $k$, \ldots. 

The further thermodynamic restrictions, arising from the vanishing of the coefficients of linear and cubic terms in the 
higher derivatives, even if their computation is straightforward, is omitted since their expression is rather long. 

Nevertheless, in Section~\ref{sec:1d}, limiting ourselves to the one-dimensional case, we solve the thermodynamic 
constraints and provide a complete physically meaningful solution.

It is easily ascertained by direct inspection of the restrictions above that the Lagrange multipliers (and 
hence the specific entropy) result to be dependent also on the gradients of the unknown variables. 

Indeed, in principle, this is also true in the classical Liu procedure, since the  Liu theorem   states that the Lagrange 
multipliers are defined on the whole state space \cite{Liu}. However, in this case, such a dependency does not 
imply an analogous dependency of the stress tensor on the gradients of the field variables. To see that, let us 
consider the case of a single non-viscous Korteweg fluid ruled by the following balance equations
\begin{equation}
\label{modelequations:sf}
\begin{aligned}
&\overline{\mathcal{E}}^{(1)}\equiv \rho_{,t}+v_j\rho_{,j}+\rho v_{j,j}=0,\\
&\overline{\mathcal{E}}^{(2)}_i\equiv \rho (v_{i,t}+v_jv_{i,j})-T_{ij,j}=0,\\
&\overline{\mathcal{E}}^{(3)}\equiv \rho (\varepsilon_{,t}+v_j\varepsilon_{,j})-T_{ij}v_{i,j}+q_{j,j}=0,
\end{aligned} 
\end{equation}
where the meaning of the symbols is analogous to the one we have adopted in the case of two fluids. Meanwhile, the 
entropy inequality takes now the form
\begin{equation}
\rho (s_{,t}+\rho v_j s_{,j}) +J_{j,j}\ge 0.
\end{equation}
Let us set the state space as
\begin{equation}
\label{statespacebis}
\mathcal{Z}=\{\rho,\varepsilon,\rho_{,j},\varepsilon_{,j},\rho_{,jk}\}.
\end{equation}
Thus, by using the classical Liu method, the entropy inequality takes the form
\begin{equation}
\rho (s_{,t}+\rho v_j s_{,j}) +J_{j,j}-\overline{\lambda}^{(1)}\overline{\mathcal{E}}^{(1)}-
\overline{\lambda}^{(2)}_i\overline{\mathcal{E}}^{(2)}_i-\overline{\lambda}^{(3)}\overline{\mathcal{E}}^{(3)}\ge 0,
\end{equation}
where $\overline{\lambda}^{(1)}$, $\overline{\lambda}^{(2)}_i$ and $\overline{\lambda}^{(3)}$ are the Lagrange 
multipliers corresponding to the local balances of mass, linear momentum and energy, respectively.
By expanding the derivatives, we get
\begin{equation}
\label{entropyconstrained:sf}
\begin{aligned}
&\left(\rho\frac{\partial s}{\partial \rho}-\overline{\lambda}^{(1)}\right)\rho_{,t}+
\rho\overline{\lambda}^{(2)}_iv_{i,t}+ \rho\left(\frac{\partial s}{\partial \varepsilon}-\overline{\lambda}^{(3)}\right)
\varepsilon_{,t}\\
&\quad+\rho \frac{\partial s}{\partial \rho_{,k}}\rho_{,kt}+  \rho\frac{\partial s}{\partial \varepsilon_{,k}}\varepsilon_{,kt} 
+\rho\frac{\partial s}{\partial \rho_{,k\ell}}\rho_{,k\ell t}\\
&\quad +\left(\rho v_j\frac{\partial s}{\partial \rho}+\frac{\partial J_j}{\partial \rho}-\overline{\lambda}^{(1)}v_j-
\overline{\lambda}^{(2)}_i\frac{\partial T_{ij}}{\partial\rho}-\overline{\lambda}^{(3)}\frac{\partial q_j}{\partial\rho}\right)
\rho_{,j}\\
&\quad +\left(\rho v_j\frac{\partial s}{\partial \varepsilon}+\frac{\partial J_j}{\partial \varepsilon}-\overline{\lambda}
^{(2)}_i\frac{\partial T_{ij}}{\partial\varepsilon}-\rho\overline{\lambda}^{(3)}v_j-\lambda^{(3)}\frac{\partial q_j}
{\partial\varepsilon}\right)\varepsilon_{,j}\\
&\quad +\left(\rho v_j\frac{\partial s}{\partial \rho_{,k}}+\frac{\partial J_j}{\partial \rho_{,k}}-\overline{\lambda}^{(2)}
_i\frac{\partial T_{ij}}{\partial\rho_{,k}}-\overline{\lambda}^{(3)}\frac{\partial q_j}{\partial\rho_{,k}}\right)\rho_{,jk}\\
&\quad +\left(\rho v_j\frac{\partial s}{\partial \varepsilon_{,k}}+\frac{\partial J_j}{\partial \varepsilon_{,k}}-
\overline{\lambda}^{(2)}_i\frac{\partial T_{ij}}{\partial\varepsilon_{,k}}-\overline{\lambda}^{(3)}\frac{\partial q_j}
{\partial\varepsilon_{,k}}\right)\varepsilon_{,jk}\\
&\quad +\left(\rho v_j\frac{\partial s}{\partial \rho_{,k\ell}}+\frac{\partial J_j}{\partial \rho_{,k\ell}}-\overline{\lambda}
^{(2)}_i\frac{\partial T_{ij}}{\partial\rho_{,k\ell}}-\overline{\lambda}^{(3)}\frac{\partial q_j}{\partial\rho_{,k\ell}}\right)
\rho_{,jk\ell}\\
&\quad-\left(\rho\overline{\lambda}^{(1)}\delta_{ij}+\rho\overline{\lambda}^{(2)}_iv_j+\overline{\lambda}^{(3)}T_{i,j}
\right)v_{i,j} \ge 0.
\end{aligned}
\end{equation}

The latter is a scalar-valued function which is linear in the 
derivatives $\rho_{,t}$, $v_{i,t}$, $\varepsilon_{,t}$, $\rho_{,kt}$, $\varepsilon_{,kt}$,
$\rho_{,k\ell t}$, $v_{i,j}$, $\varepsilon_{,jk}$ and $\rho_{,jk\ell}$ with coefficients defined on $\mathcal{Z}$ given by 
\eqref{statespacebis}.
Since these quantities are independent of the elements of the state space and can assume arbitrary values 
\cite{ColNol,Liu}, their coefficients must vanish, otherwise the above inequality could be easily violated. This leads to 
the following set of thermodynamic restrictions
\begin{equation}
\label{constraintKorteweg}
\begin{aligned}
&\frac{\partial s}{\partial\rho_{,k}}=\frac{\partial s}{\partial\varepsilon_{,k}}=\frac{\partial s}{\partial\rho_{,k\ell}}=0,\\
&\overline{\lambda}^{(1)}=\rho\frac{\partial s}{\partial\rho}=,\qquad  \overline{\lambda}^{(2)}_i=0,\qquad
\overline{\lambda}^{(3)}=\frac{\partial s}{\partial \varepsilon},\\
&\left\langle\frac{\partial J_j}{\partial \varepsilon_{,k}}-\frac{\partial s}{\partial \varepsilon}\frac{\partial q_j}
{\partial\varepsilon_{,k}}\right\rangle_{(jk)}=0,\\
&\left\langle\frac{\partial J_j}{\partial \rho_{,k\ell}}-\frac{\partial s}{\partial \varepsilon}\frac{\partial q_j}
{\partial\rho_{,k\ell}}\right\rangle_{(jk\ell)}=0,\\
&\rho^2\frac{\partial s}{\partial \rho}\delta_{i,j}+\frac{\partial s}{\partial \varepsilon}T_{ij}=0.
\end{aligned}
\end{equation}
Finally, the residual entropy inequality reduces to
\begin{equation}
\label{entropyresidual}
\begin{aligned}
&\left(\frac{\partial J_j}{\partial \rho}-\frac{\partial s}{\partial\varepsilon}\frac{\partial q_j}{\partial\rho}\right)\rho_{,j}
+\left(\frac{\partial J_j}{\partial \varepsilon}-
\frac{\partial s}{\partial\varepsilon}\frac{\partial q_j}{\partial\varepsilon}\right)\varepsilon_{,j}
+\left(\frac{\partial J_j}{\partial \rho_{,k}}-\frac{\partial s}{\partial\varepsilon}\frac{\partial q_j}{\partial\rho_{,k}}\right)
\rho_{,jk} \ge 0.
\end{aligned}
\end{equation}

Therefore, using \eqref{constraintKorteweg}, the specific entropy and the Lagrange multipliers do not depend on
the gradients entering the state space, and, in view of the last condition in \eqref{constraintKorteweg}, the Cauchy stress tensor can not depend on the 
spatial derivatives of the mass density. In other words, the classical Liu procedure leads to the conclusion that the 
Korteweg fluids do not exist in nature, since they are not in accordance with second law of thermodynamics. Of 
course, such a conclusion does not reflect the true physical reality, since Korteweg fluids exist and are involved in 
several important phenomena, such as, for instance, the capillarity phenomena. Thus, such a consequence must be 
an artifact  of the application of the classical Liu procedure. 

From the technical point of view, we observe that this result is due to the circumstance that the terms $\rho 
\frac{\partial s}{\partial \rho_{,k}}\rho_{,kt}$,
$\rho \frac{\partial s}{\partial \varepsilon_{,k}}\varepsilon_{,kt}$ and $\rho \frac{\partial s}{\partial \rho_{,k\ell}}
\rho_{,k\ell t}$ enter the entropy inequality as singletons, since the balance equations do not contain similar terms 
which can be coupled with them. Such a circumstance leads to the necessary consequence that the quantities $
\frac{\partial s}{\partial \rho_{,k}}$, $\frac{\partial s}{\partial \varepsilon_{,k}}$ and 
$\frac{\partial s}{\partial \rho_{,k\ell}}$ must vanish.

The main idea underlying the generalized Liu procedure is to create a generalized Liu inequality containing 
additional terms which can be coupled with the quantities $\rho \frac{\partial s}{\partial \rho_{,k}}\rho_{,kt}$, $\rho 
\frac{\partial s}{\partial \varepsilon_{,k}}\varepsilon_{,kt}$ and $\rho \frac{\partial s}{\partial \rho_{,k\ell}}\rho_{,k\ell t}$. 
Those terms can only be obtained by introducing into the entropy inequality the spatial gradients of the balance 
equations, up to the order of the gradients entering the state space. For instance, in the generalized entropy 
inequality \eqref{entropyconstrained}, such  terms appear due to the inclusion in the entropy inequality of the 
gradients of the balance equations. In this way, the entropy density and the Cauchy stress tensor may depend on 
the spatial gradients of the mass density and, as expected, the Korteweg fluids are fully compatible with second law 
of thermodynamics.

\section{Solution of the system of thermodynamic constraints}
\label{sec:1d}
In this Section, we prove that the constraints recovered by applying the extended procedure above described can be 
effectively solved; in fact, by considering a one-dimensional system, we proceed to the integration of the constraints 
imposed by the entropy inequality, and determine an explicit solution giving a concrete realization of the constitutive 
equations. 

The equations \eqref{modelequations} in one space dimension write in the form
\begin{equation}
\label{modelequations1d}
\begin{aligned}
&\rho_{,t}+v\rho_{,x}+\rho v_{,x}=0,\\
&\rho(c_{,t}+v c_{,x})+\left(\rho c w\right)_{,x}=0,\\
&\rho\left(v_{,t}+v v_{,x}\right)-T^{(1)}_{,x}-T^{(2)}_{,x}+\left(\frac{\rho c}{1-c}w^2\right)_{,x}=0,\\
&\rho c\left(w_{,t}+(v+w)w_{,x}+wv_{,x}\right)+(c-1)T^{(1)}_{,x}+cT^{(2)}_{,x}\\
&\qquad-c\left(\frac{\rho c}{1-c}w^2\right)_{,x}=0,\\
&\rho c(\varepsilon^{(1)}_{,t}+(v+w)\varepsilon^{(1)}_{,x})-T^{(1)}(v_{,x}+w_{,x})+q^{(1)}_{,x}=0,\\
&\rho (1-c)\left(\varepsilon^{(2)}_{,t}+
\left(v-\frac{c}{1-c}w\right)\varepsilon^{(2)}_{,x}\right)\\
&\qquad-T^{(2)}\left(v-\frac{c}{1-c}w\right)_{,x}+q^{(2)}_{,x}=0.
\end{aligned} 
\end{equation}

To proceed further, let us assume the following constitutive equations for $T^{(1)}$, $T^{(2)}$, $q^{(1)}$ 
and $q^{(2)}$:
\begin{equation}
\begin{aligned}
T^{(A)}&=\tau^{(A)}_0+\tau^{(A)}_1 \rho_{,x}^2+\tau^{(A)}_2 \rho_{,x}c_{,x}+\tau^{(A)}_3 c_{,x}^2+\tau^{(A)}
_4\rho_{,xx}+\tau^{(A)}_5 c_{,xx},\\
q^{(A)}&=q^{(A)}_1\varepsilon^{(1)}_{,x}+q^{(A)}_2\varepsilon^{(2)}_{,x},
\end{aligned}
\end{equation}
where $q^{(A)}_j$ ($j=1,2$) and $\tau^{(A)}_k$ ($k=0,\ldots,5$), with $A=1,2$, are functions assumed to depend  on 
$(\rho,c,w,\varepsilon^{(1)},\varepsilon^{(2)})$.

Moreover, let us express the specific entropy as the sum of its value at equilibrium (dependent at most on
$\rho$, $c$, $w$, $\varepsilon^{(1)}$ and $\varepsilon^{(2)}$), and a quadratic form in the gradients of the field 
variables entering the state space. 
On the basis of these assumptions, we proceed to solve the constraints arising from the entropy inequality. The lengthy
computations have been carried out by means of a package 
written by one of the authors (F. O.) in the Computer Algebra system Reduce \cite{Reduce}.

As far as the specific entropy is concerned, we obtain
\begin{equation}
\begin{aligned}
s&=\frac{1}{\rho}\left(\rho c s_{01}+\rho(1-c)s_{02}+\frac{\phi_1}{\rho c}+\phi_2\right)\\
&+\frac{\partial^3 s_1}{\partial\rho^3}\rho_{,x}^2+2 \left(\frac{\partial^2 s_2}{\partial\rho^2}-\frac{1}{\rho}\frac{\partial 
s_2}{\partial\rho}\right)\rho_{,x}c_{,x}+\rho\frac{\partial^2 s_3}{\partial\rho^2}c_{,x}^2,
\end{aligned}
\end{equation}
along with the functions $s_{01}\equiv s_{01}(\varepsilon^{(1)})$, $s_{02}\equiv s_{02}(\varepsilon^{(2)})$, 
$s_i\equiv s_i(\rho,c)$  ($i=1,2,3$), $\phi_1\equiv\phi_1(\rho c)=\phi_1(\rho^{(1)})$ and $\phi_2\equiv\phi_2(\rho(1-c))=\phi_2(\rho^{(2)})$.
In order to satisfy the principle of maximum entropy at the
equilibrium, in the expression of the entropy the quadratic part in the gradients
must be semidefinite negative, whereupon it has to be:
\begin{equation}
\begin{aligned}
&\frac{\partial^3 s_1}{\partial\rho^3}\le 0,\qquad
\rho\frac{\partial^2 s_3}{\partial\rho^2}\le 0,\\
&\rho\frac{\partial^3 s_1}{\partial\rho^3}\frac{\partial^2 s_3}{\partial\rho^2}-
\left(\frac{\partial^2 s_2}{\partial\rho^2}-\frac{1}{\rho}\frac{\partial s_2}{\partial\rho}\right)^2\ge 0.
\end{aligned}
\end{equation}

It is worth noticing that the product of the mixture density times the equilibrium part of the specific entropy is the sum 
of two contributions corresponding to the two fluids; each contribution in fact depends on the partial density and 
internal energy of the corresponding fluid. 
Let us introduce the partial temperatures of the two fluids, say
\begin{equation}
\frac{1}{\theta^{(1)}}=c \frac{d s_{01}}{d\varepsilon^{(1)}},\qquad
\frac{1}{\theta^{(2)}}=(1-c) \frac{d s_{02}}{d\varepsilon^{(2)}}.
\end{equation}

By requiring in the entropy inequality the vanishing of the coefficients of linear terms in the highest and higher 
derivatives, as well as the coefficients of the cubic terms in the higher derivatives, what remains is a
quadratic form in the gradients $(\rho_{,x},c_{,x},v_{,x},w_{,x},\varepsilon^{(1)}_{,x},\varepsilon^{(2)}_{,x})$ that must 
be positive semidefinite. The solutions of all the constraints provides the following result:
\begin{align*}
\tau^{(1)}_0&=\theta^{(1)} c\left(\phi_1^\prime-2\frac{\phi_1}{\rho c}\right),\allowdisplaybreaks\\
\tau^{(2)}_0&=\theta^{(2)}(1-c)\left(\rho(1-c)\phi_2^\prime-\phi_2\right),\allowdisplaybreaks\\
\tau^{(1)}_1&=\theta^{(1)}c\left(c(1-c)\left(\rho\frac{\partial^4 s_1}{\partial c\partial \rho^3}-\frac{\partial^3 s_1}{\partial 
\rho^2\partial c}-2\rho\frac{\partial^3 s_2}{\partial \rho^3}+2\frac{\partial^2 s_2}{\partial \rho^2}\right)-\rho^2 
c\frac{\partial^4 s_1}{\partial \rho^4}\right),\allowdisplaybreaks\\
\tau^{(2)}_1&=\frac{(c-1)\theta^{(2)}}{c\theta^{(1)}}\tau^{(1)}_1+\rho^2(1-c)\theta^{(2)}\frac{\partial^4 s_1}
{\partial\rho^4},\allowdisplaybreaks\\
\tau^{(1)}_2&=\theta^{(1)}c\left(-2\rho^2 c \frac{\partial^4 s_1 }{\partial \rho^3\partial c}
-\rho(1-2c)\frac{\partial^3s_1 }{ \partial\rho^2\partial c}+c(1-c)\frac{\partial^3 s_1 }{\partial\rho\partial c^2}\right.\\
&+\rho^2\frac{\partial^3 s_1}{\partial \rho^3}
-2\rho\frac{\partial^2 s_1}{\partial \rho^2}+2(1-2c)\frac{\partial^2s_1 }{\partial \rho\partial c}+2\frac{\partial s_1}
{\partial\rho}
+2\rho(1-c)\frac{\partial^2s_2}{\partial \rho^2}\\
&-2c(1-c)\frac{\partial^2s_2 }{\partial\rho\partial c}
-2(2-3c)\frac{\partial s_2}{\partial\rho}-2\rho^2c(1-c)\frac{\partial^3s_3 }{\partial\rho^3}\\
&\left.-2\rho c(1-c)\frac{\partial^2 s_3 }{\partial\rho^2}+2c(1-c)\frac{\partial s_3}{\partial \rho}\right),
\allowdisplaybreaks\\
\tau^{(2)}_2&=\frac{(1-c)^2\theta^{(2)}}{c^2\theta^{(1)}}\tau^{(1)}_2-2\rho(1-c)^2\theta^{(2)}\left(\rho\frac{\partial^3 
s_3}
{\partial\rho^3}+\frac{\partial^2 s_3}{\partial \rho^2}\right),\allowdisplaybreaks\\
\tau^{(1)}_3&=\theta^{(1)} c\left(\rho^2\frac{\partial^3s_1 }{\partial\rho^2\partial c}+\rho(1-2c)\frac{\partial^3s_1}
{\partial\rho\partial c^2}-c(1-c)\frac{\partial^3 s_1}{\partial c^3}-2\rho\frac{\partial^2s_1}{\partial\rho\partial c}\right.\\
&-4(1-2c)\frac{\partial^2s_1}{\partial c^2}+4\frac{\partial s_1}{\partial c}-2\rho^2 c\frac{\partial^3s_2}
{\partial\rho^2\partial c}-2\rho(1-3c)\frac{\partial^2s_2}{\partial\rho\partial c}\\
&+2c(1-c)\frac{\partial^2s_2}{\partial c^2}+2\rho\frac{\partial s_2}{\partial\rho}+8(1-2c)\frac{\partial s_2}{\partial c}+
\rho^3 c\frac{\partial^3s_3}{\partial\rho^3}\\
&-\rho^2c(1-c)\frac{\partial^3s_3}{\partial\rho^2\partial c}+\rho^2(2-c)\frac{\partial^2s_3}{\partial\rho^2}-\rho c(1-c)
\frac{\partial^2s_3}{\partial\rho\partial c}\\
&\left.-2\rho(1-2c)\frac{\partial s_3}{\partial\rho}-c(1-c)\frac{\partial s_3}{\partial c}-8s_2\right),\allowdisplaybreaks\\
\tau^{(2)}_3&=\frac{(c-1)\theta^{(2)}}{c\theta^{(1)}}\tau^{(1)}_3-\rho(1-c)\theta^{(2)}\left(
-2\rho\frac{\partial^3 s_2}{\partial\rho^2\partial c}+2\frac{\partial^2 s_2}{\partial \rho\partial c}\right.\\
&\left.+\rho^2\frac{\partial^3 s_3}{\partial\rho^3}+3\rho\frac{\partial^2 s_3}{\partial\rho^2}\right),\\
\tau^{(1)}_4&=2\theta^{(1)}c^3\left(-\rho\frac{\partial^2 s_2}{\partial\rho^2}+\frac{\partial s_2}{\partial\rho}-\rho(1-c)
\frac{\partial^2 s_3}{\partial\rho^2}\right),\allowdisplaybreaks\\
\tau^{(2)}_4&=\frac{(c-1)^3\theta^{(2)}}{c^3\theta^{(1)}}\tau^{(1)}_4+2\theta^{(2)}\rho(1-c)^3\frac{\partial^2 s_3}
{\partial\rho^2},\allowdisplaybreaks\\
\tau^{(1)}_5&=2\theta^{(1)}\rho c^2\left(-\rho\frac{\partial^2 s_2}{\partial\rho^2}+\frac{\partial s_2}{\partial\rho}-\rho(1-
c)\frac{\partial^2 s_3}{\partial\rho^2}\right),\allowdisplaybreaks\\
\tau^{(2)}_5&=\frac{(1-c)^2\theta^{(2)}}{c^2\theta^{(1)}}\tau^{(1)}_5-2\theta^{(2)}\rho^2(1-c)^2\frac{\partial^2 s_3}
{\partial\rho^2}.
\end{align*}

Furthermore, we have to impose the following inequalities:
\begin{equation}
\begin{aligned}
&s_{01}^{\prime\prime}\le 0,\qquad s_{02}^{\prime\prime}\le 0,\qquad q^{(1)}_1\le 0,\qquad q^{(2)}_2\le 0,\\
&4q^{(1)}_1q^{(2)}_2s_{01}^{\prime\prime}s_{02}^{\prime\prime}-\left(q^{(1)}_2s_{01}^{\prime\prime}+q^{(2)}
_1s_{02}^{\prime\prime}\right)^2\ge 0.
\end{aligned}
\end{equation}

Finally, the functions involved in the previous expressions must satisfy the following
differential constraints:
\begin{align*}
&c^3(1-c)^3\frac{\partial ^{4}s_{1}}{\partial c^4}-8\rho^2(7 c(1-c)-2)\frac{\partial^{2}s_{1}}{\partial \rho^2} -4 
\rho(c(11c(3-2c)-19)+4)\frac{\partial^{2}s_{1}}{\partial \rho \partial c}\allowdisplaybreaks\\
&+4c(1-c)(17c(1-c)-5)\frac{\partial^{2}s_{1}}{\partial c^2}+8\rho(7c(1-c)-2)\frac{\partial s_{1}}{\partial \rho}
\allowdisplaybreaks\\
&+20c(1-c)(1-2c)\frac{\partial s_{1}}{\partial c}+\rho^2c^2(1-c)^2\frac{\partial^{3} s_{2}}{\partial \rho^2 \partial 
c}-2c^3(1-c)^3\frac{\partial^{3} s_{2}}{\partial c^3} \allowdisplaybreaks\\
&-2\rho^2c(1-c)(1-2 c)\frac{\partial^{2} s_{2}}{\partial \rho^2}- \rho c^2(1 - c)^2\frac{\partial^{2} s_{2}}{\partial \rho 
\partial c}\allowdisplaybreaks\\
&-2\rho(1-2c)(43c(1-c)-16)\frac{\partial s_{2}}{\partial \rho}-8c(1-c)(17c(1-c)-5)\frac{\partial s_{2}}{\partial c}
\allowdisplaybreaks\\
&-\rho^3c^2(1-c)^2\frac{\partial^{3} s_{3}}{\partial \rho^3}+\rho c^3(1-c)^3\frac{\partial^{3} s_{3}}{\partial \rho \partial 
c^2}+\rho^2 c(1-c)(5c(1-c)-2)\frac{\partial^{2} s_{3}}{\partial \rho^2}\allowdisplaybreaks\\
&+c^3(1-c)^3\frac{\partial^{2} s_{3}}{\partial c^2}+12\rho c(1-c)(7c(1-c)-2)\frac{\partial s_{3}}{\partial \rho}
\allowdisplaybreaks\\
&-4c^2(1-c)^2(1-2c)\frac{\partial s_{3}}{\partial c}-40c(1-c)(1-2c)s_{2}=0,\allowdisplaybreaks\\
&\rho^2\frac{\partial^{3}s_{1}}{\partial \rho^3}+\rho(1-2c)\frac{\partial^{2}s_{2}}{\partial \rho^2}-(1-2c)\frac{\partial 
s_{2}}{\partial \rho}-\rho c(1-c)\frac{\partial^{2}s_{3}}{\partial \rho^2}=0,\allowdisplaybreaks\\
&\rho\left(\frac{\partial ^{3}s_{1}}{\partial \rho^{2}\partial c}-\frac{\partial^{2}s_{2}}{\partial \rho^{2}}\right)-\frac{\partial 
s_{2}}{\partial \rho}=0,\allowdisplaybreaks\\
&c(1-c)\frac{\partial^{3}s_{1}}{ \partial \rho\partial c^2}-2\rho\frac{\partial^{2}s_{1}}{ \partial \rho^2}+2(1-2c)
\frac{\partial^{2}s_{1}}{ \partial \rho\partial c}+2\frac{\partial s_{1}}{ \partial \rho}-2c(1-c)\frac{\partial^{2} s_{2}}{ \partial 
\rho \partial c}\allowdisplaybreaks\\
&-4(1-2c)\frac{\partial s_{2}}{ \partial \rho}+\rho c(1-c)\frac{\partial^{2} s_{3}}{ \partial \rho^2}+2c(1-c)\frac{\partial 
s_{3}}{ \partial \rho}=0,\allowdisplaybreaks\\
&c^2(1-c)^2\frac{\partial^{3}s_{1}}{ \partial c^3}-2\rho^2(1-2c)\frac{\partial^{2}s_{1}}{ \partial \rho^2}-2\rho(3c(1-
c)-1)\frac{\partial^{2}s_{1}}{ \partial \rho\partial c}\allowdisplaybreaks\\
&+4c(1-c) (1-2c)\frac{\partial^{2}s_{1}}{ \partial c^2}+2\rho(1-2c)\frac{\partial s_{1}}{ \partial \rho}-4c(1-c)\frac{\partial 
s_{1}}{ \partial c}\allowdisplaybreaks\\
&-2c^2(1-c)^2\frac{\partial^{2}s_{2}}{ \partial c^2}+4\rho(3c(1-c)-1)\frac{\partial s_{2}}{ \partial \rho}-8c(1-c)(1-2c)
\frac{\partial s_{2}}{ \partial c}\allowdisplaybreaks\\
&+\rho c^2(1-c)^2\frac{\partial^{2}s_{3}}{ \partial\rho\partial c}+4\rho c(1-c)(1-2c)\frac{\partial s_{3}}{ \partial\rho}
+c^2(1-c)^2\frac{\partial s_{3}}{ \partial c}+8c(1-c)s_{2}=0.
\end{align*}

Also, an expression for the entropy flux, made by the classical expression, an additional term proportional to the 
diffusion velocity $w$, and a quadratic form in the gradients $\rho_{,x}$, 
$c_{,x}$, $v_{,x}$ and $w_{,x}$, is found:
\begin{equation}
\begin{aligned}
J_s&=\frac{q_1}{c\theta^{(1)}}+\frac{q_2}{(1-c)\theta^{(2)}}
+\left(\rho c(s_{01}-s_{02})+\frac{\phi_1}{\rho c} -\frac{c}{1-c}\phi_2\right)w\\
&+\frac{c w}{\rho}\left(\rho\frac{\partial^{2} s_{2}}{\partial \rho^2}
-\frac{\partial s_{2}}{\partial \rho}\right)\rho_{,x}^{2}+2w\left(\rho\frac{\partial^{2} s_{2}}{\partial \rho^2}
-\frac{\partial s_{2}}{\partial \rho}\right)\rho_{,x}c_{,x}\\
&-2\left((1-2c)\left(\rho\frac{\partial^{2} s_{2}}{\partial \rho^2}
-\frac{\partial s_{2}}{\partial \rho}\right)-\rho c(1-c)\frac{\partial^{2} s_{3}}{\partial \rho^2}\right)\rho_{,x}v_{,x}\\
&+2c\left(\rho\frac{\partial^{2} s_{2}}{\partial \rho^2}
-\frac{\partial s_{2}}{\partial \rho}\right)\rho_{,x}w_{,x}-\frac{\rho w}{1-c}\left(\rho\frac{\partial^{2} s_{2}}{\partial \rho^2}
-\frac{\partial s_{2}}{\partial \rho}-\rho\frac{\partial^{2} s_{3}}{\partial \rho^2}\right)c_{,x}^{2}\\
&+2\rho\left(\rho\frac{\partial^{2} s_{2}}{\partial \rho^2}
-\frac{\partial s_{2}}{\partial \rho}\right)c_{,x}v_{,x}+2\rho^2 c\frac{\partial^{2} s_{3}}{\partial \rho^2}c_{,x}w_{,x}.
\end{aligned}
\end{equation}

Thus, the results above detailed complete the exploitation of entropy inequality 
for the one-dimensional case of a mixture of two Korteweg fluids with two velocities and two temperatures.

\section{Discussion}
\label{sec:final}
In this paper, we have analyzed the thermodynamics of binary mixtures with two temperatures and two velocities, a 
topic which has been scarcely considered heretofore in the literature. Some technical problems connected with our approach 
have been faced.
First, the correct formulation of the partial balances for each constituent, by taking into account that, since these 
constituents have different velocities, the material time derivative for each of them has a different definition.
Then, we were able to prove, by means of an extended Liu procedure, that the strong nonlocality of the constitutive 
equations for the stress tensors of the constituents is compatible with second law of thermodynamics. As a result, 
the thermodynamic compatibility of the constitutive equation of the stress tensor which is sufficiently general to 
include that of classical Korteweg fluids, proposed by D.~J.~Korteweg in 1901, has been proved. 
A complete solution of the system of the thermodynamic restrictions has been computed by means of a package 
written by one of the authors (F.~O.) in the Computer Algebra System Reduce \cite{Reduce}. Although this solution 
regards a one-dimensional system, it is complete, physically sound and immediately applicable. 
Work is in progress to provide a solution of the constraints imposed by second law of thermodynamics in two and 
three space dimensions. Such solutions will allow us to study the form of the phase boundaries at the equilibrium of 
phases on a purely mechanical basis. This is a further difficult and scarcely explored problem. From the 
mathematical point of view, for a single non-viscous Korteweg fluid it consists in finding the solutions of the following 
problem
\begin{equation} 
\label{korteq}
\begin{aligned}
\frac{\partial}{\partial x_j} &\left(\left(-p+\sum_{k=1}^3\left(\alpha_1\frac{\partial^2\rho}{\partial x_k^2}+
\alpha_2\frac{\partial\rho}{\partial x_k}\frac{\partial\rho}{\partial x_k}\right)\right)\delta_{ij}\right.\\
&\left.+\alpha_3\frac{\partial\rho}{\partial x_i}
\frac{\partial\rho}{\partial x_j}+\alpha_4\frac{\partial^2\rho}{\partial x_i\partial x_j}\right)=0.
\end{aligned}
\end{equation}
Remarkably, in 1983, Serrin \cite{Serrin} proved that, unless rather special conditions are satisfied, the only 
geometric phase boundaries which are consistent with equation~\eqref{korteq} are either spherical, cylindrical, or 
planar.
More in detail, based on a general theorem proved in \cite{Pucci}, Serrin was able to prove that the constitutive 
equation \eqref{kort} must be such that the coefficients therein involved satisfy the condition
\begin{equation}
\label{condkort}
A \equiv b c+\frac{1}{2}\left(c^2-a\frac{\partial c}{\partial \rho}\right)=0,
\end{equation}
where
\begin{equation}
a=\alpha_1+\alpha_4,\qquad b=\alpha_2+\alpha_3,\qquad c=\frac{\partial\alpha_4}{\partial\rho}-\alpha_3.
\end{equation}

From the mathematical point of view, this constraint can  be understood as a consequence of the fact that the 
equilibrium equation  \eqref{korteq} has three independent components while liquid-vapor phase equilibria are 
determined by just one physical variable, namely the mass density,  \emph{i.e}, the equilibrium system \eqref{korteq} 
is overdetermined.  In this situation, Pucci  \cite{Pucci} has shown that any solution of this system necessarily has 
only level surfaces with constant mean and Gaussian curvature \cite{MOV}, 
which are either (pieces of) concentric spheres, or concentric circular cylinders, or parallel planes.

From the physical point of view, this  result reflects the experimental evidence that several (but not all) phase 
boundaries  have constant mean curvature.
On the other hand, the question can be raised whether equation \eqref{condkort} is  physically necessary, since 
without it the theory allows very few equilibrium configurations. It is worth observing that, although rather unusual, 
this fact  surely does not necessarily lead to the condition $A = 0$.  Moreover, as it turns out by the results of Section 
\ref{sec:liu},  the second law of thermodynamics, in the form of the generalized Clausius-Duhem inequality 
\eqref{entropyconstrained}, does not require $A = 0$. 
Under these circumstances, Serrin argued that his result only offers significant  reasons to accept the restriction $A = 
0$ for any physically realistic Korteweg fluid \cite{Serrin}, although it is not a necessary condition for the equilibrium. 

It is worth observing that for the solution provided in Section~\ref{sec:1d} we do not have an overdetermined system, 
since the equilibrium conditions are expresses by the two equations
\begin{equation}
\frac {\partial}{\partial x} \left(\tau^{(A)}_0+\tau^{(A)}_1 \rho_{,x}^2+\tau^{(A)}_2 \rho_{,x}c_{,x}+\tau^{(A)}_3 c_{,x}^2+
\tau^{(A)}_4\rho_{,xx}+\tau^{(A)}_5 c_{,xx}\right)=0,
\end{equation}
with $A=1,2$, in the two unknown functions $\rho$ and $c$. Moreover, for the one-dimensional system at hand, the 
phase boundaries reduce always to a single point. 

A different situation could arise in the multi-dimensional cases, in which the equilibrium of two different fluids 
could require even more severe constraints with respect to Equation~\eqref{condkort}. 
In these latter cases, we have to exploit  a similar, and probably more complicated, analysis along the lines of 
\cite{Serrin} in order to determine possible additional compatibility conditions on the coefficients of the Cauchy stress 
tensors. Such a problem deserves consideration, and will be the subject of  a forthcoming paper.

Further research will also concern the investigation of the possibility of applying the present results to liquid 
Helium 2 below the so called $\lambda$-point. In fact, it is well known that, in this condition, Helium 2 behaves as a mixture of 
a fluid component and a superfluid one \cite{Khal}. In such a situation nonlocal effects are detectable and nonlocal 
constitutive equations are important \cite{AtFox}. The interesting fields are the partial mass densities and the two 
velocities \cite{AtFox}, Thus, it could be interesting to investigate if two different temperatures could play any role 
and could provide a more deep insight in the physics of the problem.

\bigskip
\noindent
\textbf{Acknowledgments.}
Work supported by the ``Gruppo Nazionale per la Fisica Matematica'' (GNFM) of the Istituto Nazionale di Alta 
Matematica ``F. Severi''.

\end{document}